\documentclass[onecolumn,showpacs,preprintnumbers,amsmath,amssymb]{revtex4}
\usepackage{graphicx}
\usepackage{dcolumn}
\usepackage{bm}
\usepackage{color}
\usepackage{amsmath}
\usepackage{xcolor}
\usepackage[normalem]{ulem}
\usepackage{epstopdf}

\newcommand{\ns}{\,ns^{-1}}

\begin{document}

\title{Propagation studies for the construction of atomic macro-coherence in dense media as a tool to investigate neutrino physics}

\author{J. Mart\'in Vaquero\,$^1$}
\author{J. Cuevas--Maraver\,$^2$}
\author{A. Peralta Conde\,$^3$}
\affiliation{$^1$ Facultad de Ciencias. Universidad de Salamanca. 37008 Salamanca, Spain.\\
$^2$ Grupo de F\'{i}sica No Lineal, Departamento de F\'{i}sica Aplicada I,
Universidad de Sevilla. Escuela Polit\'{e}cnica Superior, C/ Virgen de \'{A}frica, 7, 41011-Sevilla, Spain \\
Instituto de Matem\'{a}ticas de la Universidad de Sevilla (IMUS). Edificio Celestino Mutis. Avda. Reina Mercedes s/n, 41012-Sevilla, Spain. \\
$^3$ Centro de L\'aseres Pulsados, CLPU, Parque Cient\'ifico, 37185 Villamayor, Salamanca, Spain.}

\begin{abstract}

In this manuscript we review the possibility of inducing large coherence in a macroscopic dense target by using adiabatic techniques. For this purpose we investigate the degradation of the laser pulse through propagation, which was also related to the size of the prepared medium. Our results show that, although adiabatic techniques offer the best alternative in terms of stability against experimental parameters, for very dense media it is necessary to engineer laser-matter interaction in order to minimize laser field degradation. This work has been triggered by the proposal of a new technique, namely Radiative Emission of Neutrino Pairs (RENP), capable of investigating neutrino physics through quantum optics concepts which require the preparation of a macrocoherent state. 

\end{abstract}


\maketitle

\section{Introduction} \label{Introduction}

We are far from understanding neutrinos despite our enormous theoretical and experimental efforts. Grasping the physics of these elusive particles is crucial because of important unanswered questions and weaknesses of available theories. For example we do believe that neutrinos are responsible for the matter-antimatter asymmetry of the Universe but we still do not have experimental proof of it. At present, two major questions about neutrinos can be posed, namely their nature and their mass hierarchy, although surely more questions will rise as soon as we deepen our knowledge of neutrino physics.

As to the nature of neutrinos, several scientific collaborations run nowadays aiming to determine whether neutrinos are Majorana or Dirac particles. That means, whether neutrino and antineutrino are the same particle or different ones. We are currently testing different technologies in experiments like: GERDA based on high resolution calorimeters with high purity enriched $^{76}$Ge crystal diodes experiment located at the Laboratori Nazionali del Gran Sasso (LNGS, Italy) (see for example \cite{GERDA13} and references therein), EXO-200 based on isotopically enriched liquid $^{136}$Xe located  at the Waste Isolation Pilot Plant (WIPP) in New Mexico, USA \cite{EXO_2}, or KamLAND-Zen in Japan \cite{KamLAND} and NEXT at the Canfranc Underground Laboratory in Spain \cite{NEXT12} based on ultrapure high pressure $^{136}$Xe.

The neutrino hierarchy is also subject of theoretical and experimental investigation by several scientific collaborations. Experiments like T2K (Tokai to Kamioka) in Japan (see for example \cite{T2K16} and references therein), NO$\nu$A (NuMI Off-Axis $\rm \nu_e$ Appearance) in USA (see \cite{NOvA16} and reference therein), and LNBE (long-baseline neutrino experiment) also in USA \cite{LNBE} rely on the interaction of neutrinos with matter as they pass through Earth. Since this interaction is extremely weak, these experiments require extremely large, complex and expensive scientific installations. However, a recent proposal by Yoshimura and colleagues from the University of Okayama in Japan may revolutionize our conception of neutrino experiments (see \cite{Fukumi12} and references therein). This new approach relies on the synergy of different scientific fields, like particle Physics, laser Physics and quantum optics, to investigate neutrino nature and hierarchy with much simpler (and cheaper) experimental setups.

The key idea of this ambitious proposal is to induce the collective de-excitation of a previously prepared medium in a coherent superposition of states. Provided that the upper state is metastable, the standard electroweak theory allows a relaxation path ---known as Radiative Emission Neutrino Pair (RENP)--- consisting in the emission of a neutrino pair and a photon \cite{Aitchison03}. Since the neutrinos and the photon originate from the same process, by analyzing the energy spectrum of the photon we may infer the individual neutrino masses and hence the neutrino hierarchy. Also RENP could potentially elucidate if neutrinos are Majorana or Dirac particles.

Although RENP is theoretically feasible, the realization of experiments is on the edge of our technical capabilities. The extremely small cross section of the process makes the photon analysis extremely challenging, being necessary lives times of the order of days to years \cite{Song16}.  However on one hand we can increase the signal yield by previously preparing the target in a maximum superposition of states ---similarly to the process of superradiance suggested by Dicke in 1954 \cite{Dicke54}--- where the signal yield rather than scaling with N, which is clearly insufficient, scales with N$^2$ being N the number of particles of the target. On the other hand, and as a previous step to RENP, we must develop methods and techniques that provide us with an exquisite control over the amplification mechanisms in previously prepared macroscopic targets \cite{Miyamoto14, Miyamoto15, Masuda15}. According to this, it seems mandatory to develop robust laser-matter interaction techniques capable of inducing maximum coherence in macroscopic targets.

The original proposal from Yoshimura and coworkers was to use a Raman process via virtual states to prepare the medium \cite{Sokolov97}. Using this technique Miyamoto \emph{et al} reported an induced coherence of $\sim6.5\,\%$ that is far from optimum \cite{Miyamoto14, Miyamoto15, Masuda15}. The main reason of this low induced coherence is that Raman processes are not robust enough with respect to experimental parameters like e.g. laser energy fluctuations. Motivated by this problem and based on the concept of adiabatic evolution (see for example \cite{Shore08}), we presented recently a new adiabatic technique called two-photon Coherent Population Return (CPR) capable of inducing a robust coherence close to 100\,\% \cite{Boyero15}.

In this article, we explore the possibility of inducing maximum coherence in a macroscopic target using two-photon CPR. For this task we will study in detail the propagation of the radiation fields through the medium using the Maxwell-Bloch equations as a function of the medium density, paying special attention to those parameters relevant for an experimental implementation of the technique, as is the spatial distribution of intensities across the laser profile. We also investigate the amplification properties of such prepared medium for externally triggered two-photon coherent emission. As we will describe later, the observation and control of this emission ---which is an unavoidable source of background--- is a required step for the observation of the much weaker RENP process.

Finally, we would like to note that although our work has been motivated by RENP and our discussion focuses on this topic, the preparation of quantum superposition of states, and therefore the maximization of the nonlinear response of a medium, is an active field of research with applications, e.g., in high harmonic generation \cite{Chacon15} or the production of short extreme-ultraviolet radiation \cite{Halfmann10}. Thus, the results of this manuscript can be directly applied to other different fields of science.

\section{Two-photon CPR}

\begin{figure}
\includegraphics[width=0.5\columnwidth]{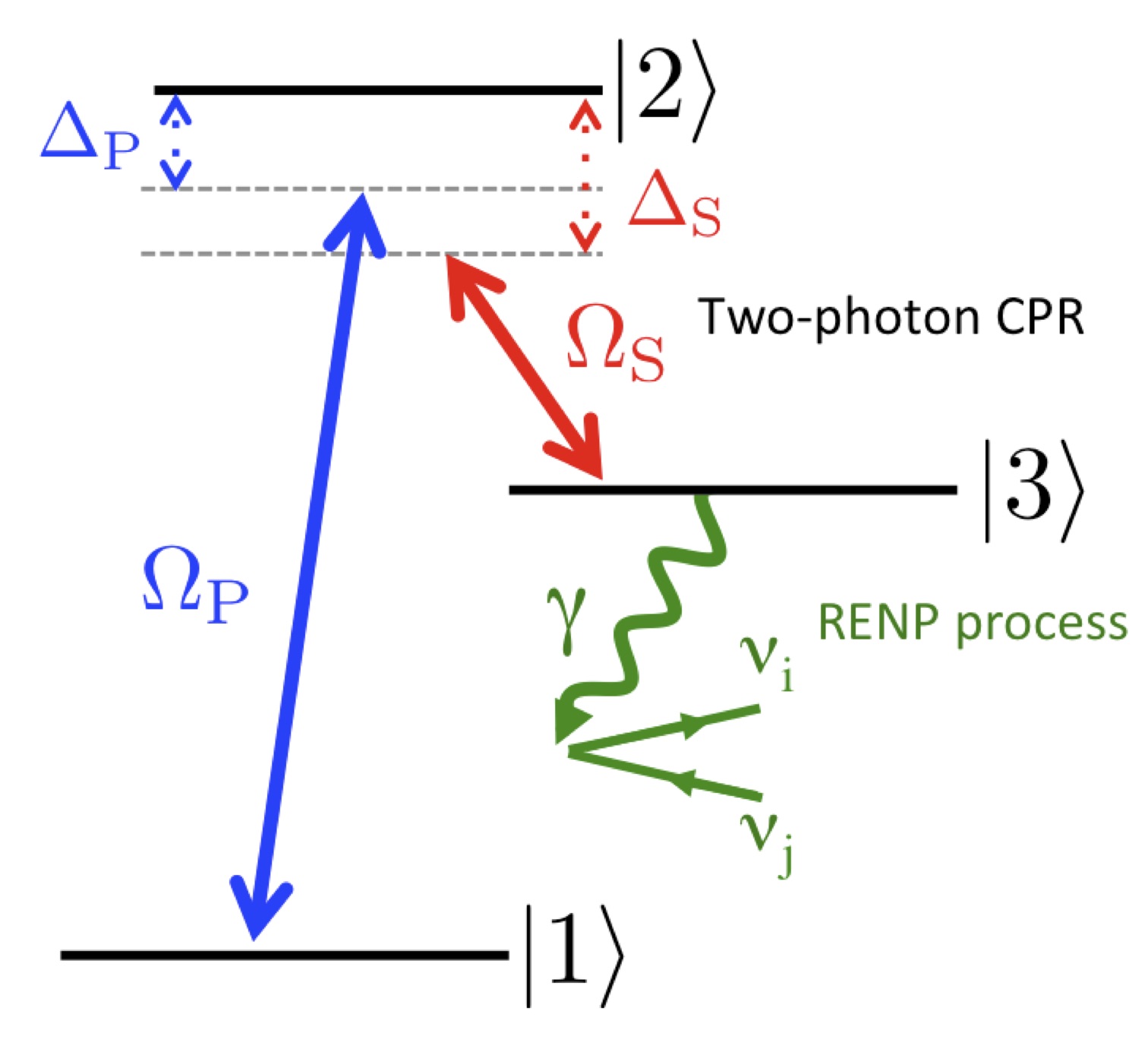}
\caption{\label{fig1} (Color online) Level scheme.}
\end{figure}

Figure\,\ref{fig1} shows the system of interest for the investigation of the neutrino hierarchy using RENP. In this system, usually referred in the literature as a $\Lambda$-system, the transitions $|1\rangle\leftrightarrow|2\rangle$ and $|2\rangle\leftrightarrow|3\rangle$ are electric dipole allowed, and consequently the state $|3\rangle$ is a metastable one. The key aspect of RENP is the construction of a robust coherence between the states $|1\rangle$ and $|3\rangle$ that may amplify the weak relaxation path of the latter state, consisting on the emission of a photon and two neutrinos.

\begin{figure}
\includegraphics[width=0.5\columnwidth]{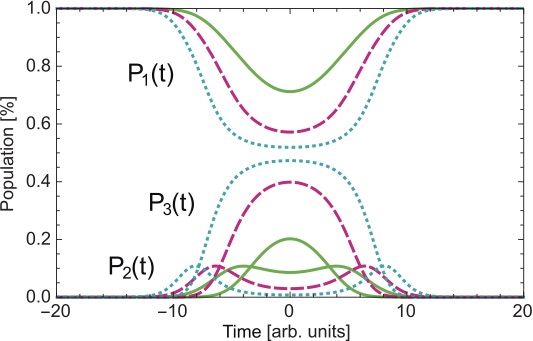}
\caption{\label{fig2} (Color online) Population of the different states as a function of time for different $\rm \Omega_{0p}/\Delta_P$ ratios. The Rabi frequencies are $\rm \Omega_P=\Omega_S=\Omega_{0p}\exp{(-t^2/\tau^2)}$ with $\tau=6$ \,a.u., and $\rm \Delta_S=2\Delta_P$. The solid green line corresponds to a $\rm \Omega_{0p}/\Delta_P$ ratio of 2, the dashed pink line to $\rm \Omega_{0p}/\Delta_P=4$, and the dotted blue line to $\rm \Omega_{0p}/\Delta_P=10$.}
\end{figure}

As we studied in \cite{Boyero15}, the parameter region that produces a robust coherence $\rm \rho_{13}$ in the medium is defined by the following conditions $\rm \frac{1}{\tau_p} <\Delta_P<\Delta_S$ and $\rm \Omega_P\simeq\Omega_S$ being $\rm \tau_P$ the pulse duration of the Pump laser, $\rm \Omega_i$ the Rabi frequencies defined as $\rm \Omega_i=\mu_iE_i/\hbar$ where $\rm \mu_i$ is the dipole transition moment and $\rm E_i$ the electric field of the laser, and $\rm \Delta_i$ the detuning with respect to the resonance. Provided that these conditions are fulfilled and the interaction is strong enough, the populations transferred to states $|1\rangle$ and $|3\rangle$ approach $50\%$ and, hence, a maximum coherence, i.e., $\rm |\rho_{13}|=\frac{1}{2}$, is built in the system (see Fig.\,\ref{fig2}). We have assigned $\rm |\rho_{13}|=\frac{1}{2}$ to 100\,\%.

It is important to note that unlike other techniques for achieving coherence in a medium (see for example \cite{Shore99}) two-photon CPR builds a transient coherence vanishing once the interaction with the Pump and Stokes lasers cease. This is not a major constraint if the process of interest is temporally of the order of this transient coherence.

\section{Wave propagation}

\subsection{The model}

In order to study the propagation effects of electromagnetic radiation through a macroscopic medium, it is necessary to solve the Maxwell-Bloch equations. In this formalism, Liouville equation rules the population dynamics of the system, and Maxwell equations the propagation effect, being the coherence between the states of the system the link between both sets of equations. Explicitly, the equation that governs the system evolution reads
\begin{equation}\label{Liouville}
\rm \imath \hbar \frac{d\rho}{d t}=\left[H, \rho \right]
\end{equation}
being $\rho$ the density matrix
\begin{equation}
\rm \rho=\left(
 \begin{array}{ccc}
\rho_{11} & \rho_{12} & \rho_{13} \\
\rho_{21} & \rho_{22} & \rho_{23} \\
\rho_{31} & \rho_{32} & \rho_{33}
 \end{array}
\right).
\end{equation}
where the diagonal elements $\rm \rho_{ii}$ represent the populations of the state $\rm |i\rangle$, the non-diagonal elements the coherences, and H the hamiltonian of the system described in Fig.\,\ref{fig1} which reads in the Rotating Wave Approximation (RWA) as
\begin{equation}
\label{Hamilt}
\rm H=\frac{\hbar}{2}\left(
\begin{array}{ccc}
0 & \rm \Omega_P  & 0\\
\rm \Omega_P  & \rm 2\Delta_P & \rm \Omega_S  \\
\rm 0 & \rm \Omega_S  &\rm 2(\Delta_P-\Delta_S)
\end{array}
\right).
\end{equation}
Assuming the Slow Varying Envelope approximation (SVEA) and a coordinate system that moves parallel to the laser pulses, the equations for the envelope variation of the Pump and Stokes lasers read \cite{Sh90}
\begin{eqnarray}
\label{propag}
\rm
\frac{\partial \Omega_P(z, t)}{\partial z}=-\frac{N k_P \mu_{12}^2}{\hbar\epsilon_0}\Im[\rho_{12}(z, t)] \nonumber \\
\rm
\frac{\partial \Omega_S(z, t)}{\partial z}=-\frac{N k_S \mu_{23}^2}{\hbar\epsilon_0}\Im[\rho_{23}(z, t)]
\end{eqnarray}
where N is the particle density, k$\rm _i$ the wavenumber, and $\rm \mu_{ij}$ the transition dipole moment.

\subsection{Numerical results}

\begin{figure}
\includegraphics[width=0.5\columnwidth]{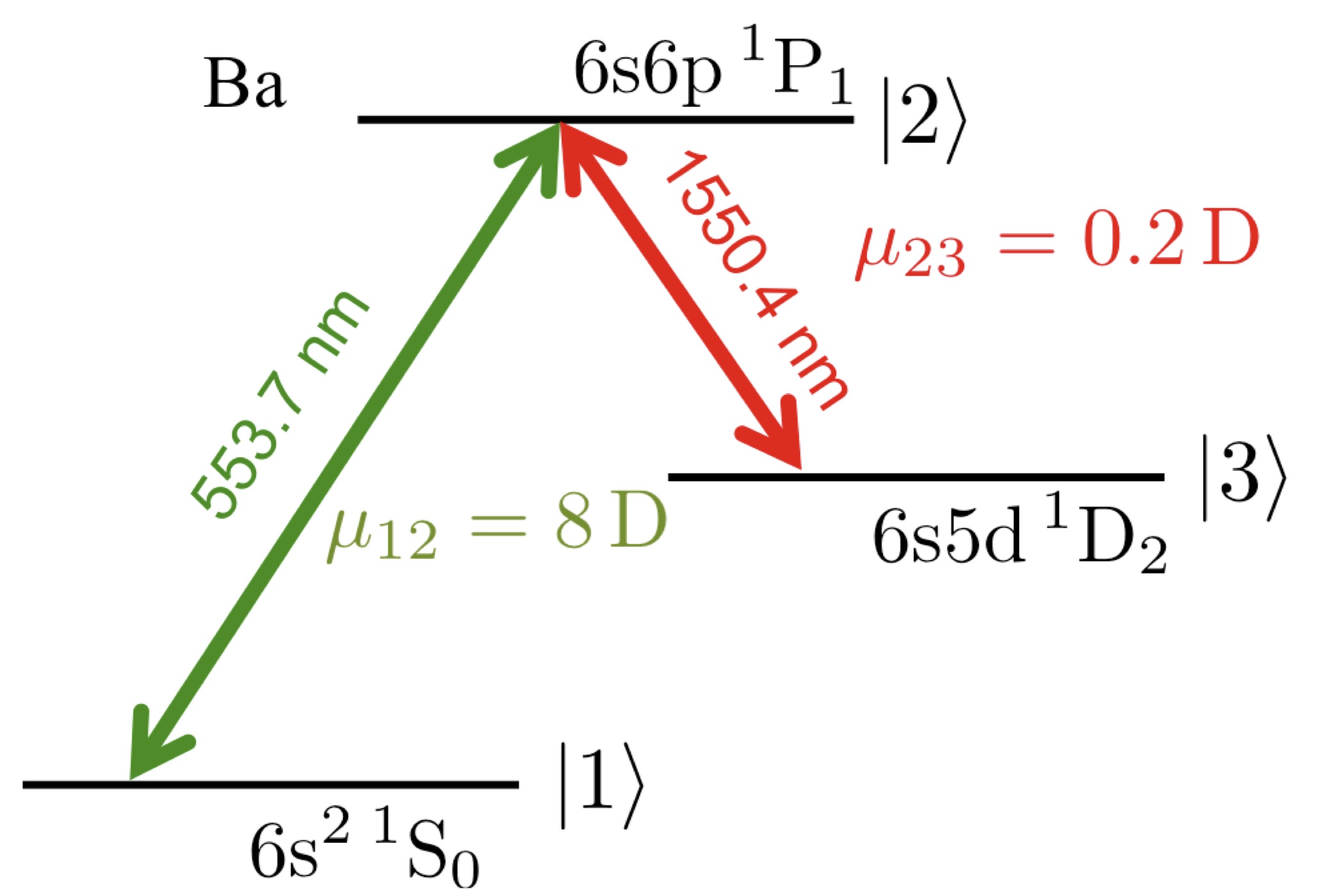}
\caption{\label{fig3} (Color online) Level scheme for barium with the corresponding dipole transition moments (1 Debye~\,=~\,$3.33\times10^{-30}$\,C$\cdot$m).}
\end{figure}

In order to numerically analyze light propagation in a given medium, we need to use the level scheme of the corresponding material. In our case, we have chosen barium (one of the possible candidates for RENP \cite{Fukumi12}) as this material. Its level scheme is shown in Fig.\,\ref{fig3}.

The numerical scheme used for solving the Maxwell-Bloch equations \ref{Liouville} and \ref{propag} consists basically in following the two-step iterative procedure: first of all, Liouville equation (\ref{Liouville}) is numerically integrated from $\rm t=-100$\,a.u. to $\rm t=100$\,a.u. so that the temporal profiles of the Pump and Stokes pulses are introduced in the hamiltonian (\ref{Hamilt}); in the first step, i.e., at $\rm z=0$, these profiles are given by $\rm \Omega_P(t)=\Omega_{0P}Exp[-4\ln 2 \frac{t^2}{\tau_P^2}]$, $\rm \Omega_S(t)=\Omega_{0P}Exp[-4\ln 2 \frac{t^2}{\tau_S^2}]$. Once the density matrix is found at every prescribed value of t, the propagation of the laser pulses can be attained by solving Eq.\,(\ref{propag}) mapping the laser profile for all t at a given propagation step. The iteration procedure is restarted by introducing the propagated laser profile at $\rm \Omega_i(z+\delta z, t)$ into the Liouville equation.

\begin{figure}
\includegraphics[width=0.5\columnwidth]{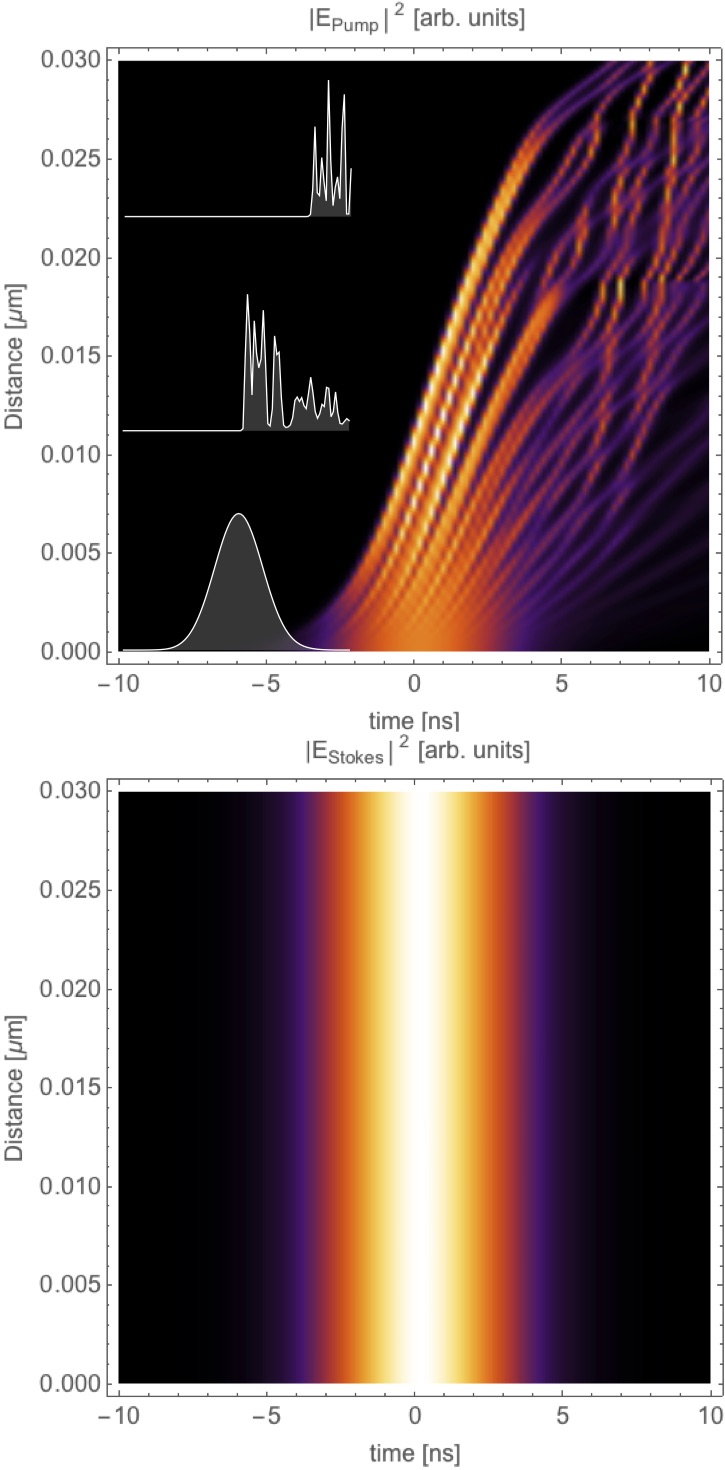}
\caption{\label{fig4} (Color online) Degradation of Pump and Stokes pulses as they propagate through a macroscopic target. The simulation data were: $\rm \Omega_i(t,z=0)=\Omega_{0i}\exp[-4\ln 2 \frac{t^2}{\tau_i^2}]$ with $\rm\Omega_{0P}=\Omega_{0S}=20\ns$ and $\rm \tau_P=\tau_S=7\,ns$, $\rm\Delta_P=4\ns$, $\rm \Delta_S=9\ns$, and N=10$^{7}$\,part./$\mu$m$^3$. The step size of the propagation equations was $\rm \delta z=10^{-6}\,\mu$m.}
\end{figure}

Figure\,\ref{fig4} shows the degradation of the Pump and Stokes lasers as they propagate through the target. For this special case of Ba, the Pump pulse degrades rapidly while the Stokes pulse remains almost undeformed for the same traveled distance, which is caused by the difference in dipole transition momenta. The peculiar deformation of the Pump pulse has its origin in the population dynamics depicted in Fig.\,\ref{fig2}. At the beginning of the interaction the pulse gives energy to the medium to produce the population excitation, but once the interaction ceases and all the population returns to the ground state, the medium returns this energy. As a consequence, the rising part of the Pump pulse gets sharper  when propagating.

\begin{figure}
\includegraphics[width=0.5\columnwidth]{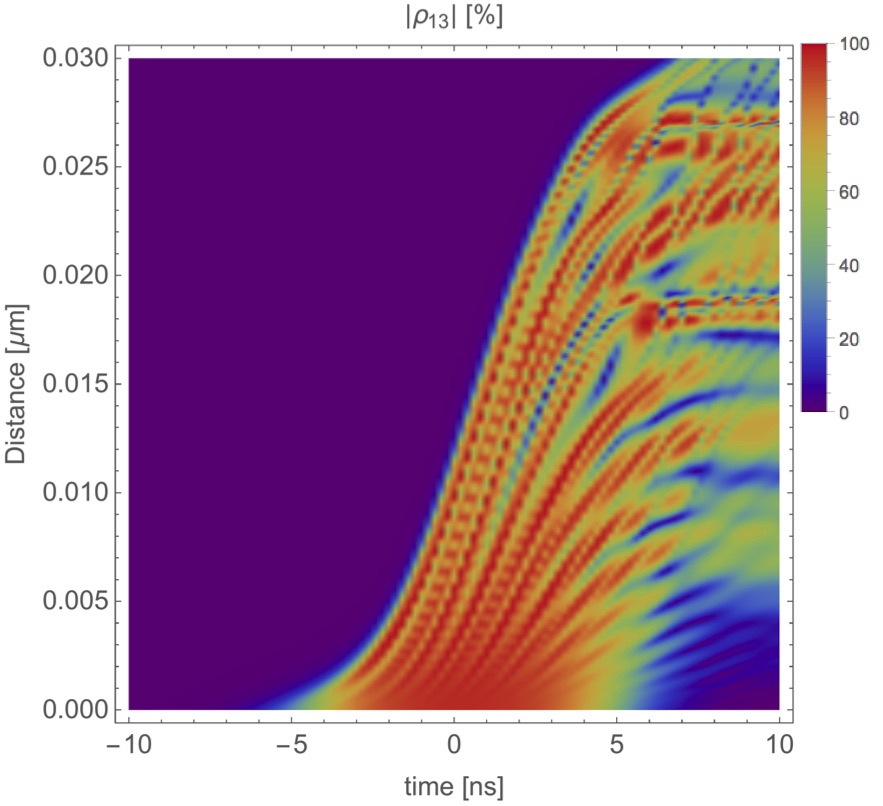}
\caption{\label{fig5} (Color online) Coherence $\rm |\rho_{13}|$ as a function of time and space for the same simulation parameters of Fig.\,\ref{fig4}. We refer $\rm |\rho_{13}|=1/2$ to 100\%.}
\end{figure}

However, despite the noticeable effect that the interaction with the medium produces over the laser pulses, the robustness of two-photon CPR allows to construct a wide ---spatially speaking--- coherence in the target. Figure\,\ref{fig5} shows the coherence  $\rm |\rho_{13}|$ as a function of time and space. The region where the coherence is close to 100\%, although transient, is sufficiently large for offering good perspectives on the observation and control of coherent two-photon emission, and hence for a potential application to RENP.

\subsection{Externally triggered coherent two-photon emission}

\begin{figure}
\includegraphics[width=0.5\columnwidth]{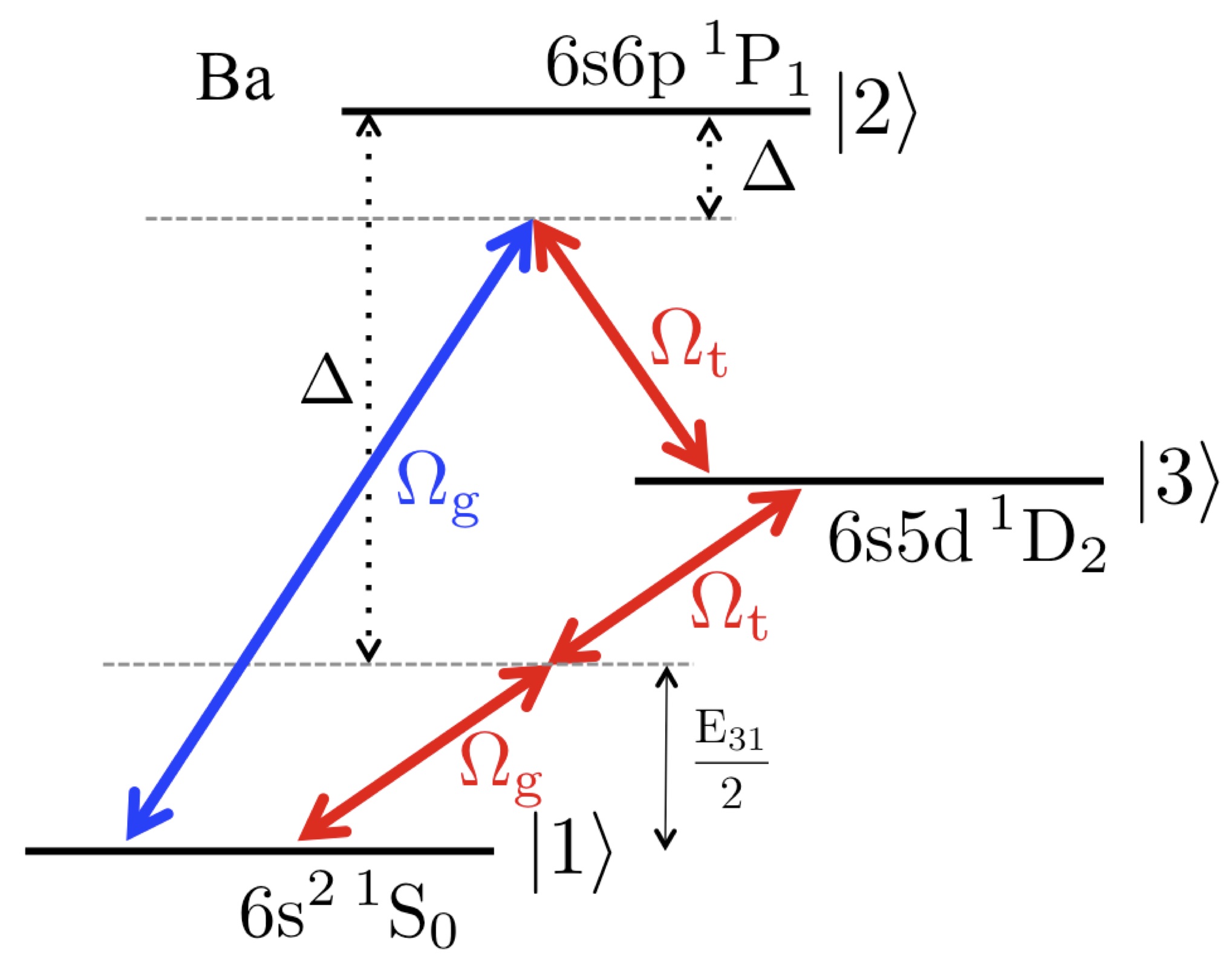}
\caption{\label{fig6} (Color online) Level scheme for trigger and generated fields.}
\end{figure}

The basic process of RENP is the atomic transition $\rm |3\rangle\rightarrow|1\rangle+\gamma+\nu_i\bar{\nu}_j$ via an intermediate virtual state $\rm |p\rangle$ with $\rm E_p>E_3>E_1$, being the transition $\rm |p\rangle\leftrightarrow|1\rangle$ of type E1 leading to the photon emission, and the transition $\rm |p\rangle\leftrightarrow|3\rangle$ of type M1 leading to the emission of the neutrino pair. This relaxation path is stimulated by two trigger lasers of frequencies $\rm \omega$ and $\rm \omega'$ with energies defined by $\rm \hbar\omega+\hbar\omega'=E_{31}$. Every time a pair of massive neutrino species is emitted, the emitted photon energy decreases by

\begin{equation}\label{neutrin_diff}
   \rm E_\gamma=\frac{E_{31}}{2}-\frac{(m_i+m_j)^2}{2E_{31}}
\end{equation}

with $\rm m_i$ being the mass of the three massive neutrino states that combined in linear quantum superposition give rise to the three known neutrinos $\rm \left(\nu_e, \nu_\mu, \nu_\tau\right)$ \cite{Fukumi12, Nota}. Thus, scanning the trigger frequencies around the different threshold positions and analyzing the spectrum of the emitted photon, it is possible to obtain information about the neutrino masses \cite{Fukumi12, Song16}. Unfortunately this process is not the only one taking place in the system, being a major source of background the two-photon coherent emission via the dipole allowed state $\rm |2\rangle$ with energy $\rm E_{31}/2$ shown in Fig.\,\ref{fig6}. Although the ``upper path'' is considerably more intense than the lower one ---the generated field is inversely proportional to the detuning as it is indicated in Eqs.\,\ref{gener} and \ref{gener1} below---, this low energy path represents a critical source of noise for the observation of RENP. Accordingly, it is a necessary step {\em en route} to controlling in detail the coherent enhancement of the externally triggered two-photon emission.

Assuming a previously prepared medium in a coherent state, and a trigger laser described by its Rabi frequency $\rm \Omega_t(z, t)$ the coupled set of differential equations that described the evolution of the trigger pulse and the coherent two-photon emission (namely, the generated field) $\rm \Omega_g(z, t)$ are \cite{Harris96}:
\begin{eqnarray}
\label{gener}
\rm \frac{d\Omega_t(z, t)}{dz}=-\imath\xi_t\rho_{31}(t)\Omega_g(z, t) \nonumber \\
\rm \frac{d\Omega_g(z, t)}{dz}=-\imath\xi_g\rho_{13}(t)\Omega_t(z, t)
\end{eqnarray}
being
\begin{equation}
\label{gener1}
\rm \xi_t=\frac{k_tN\mu_{23}^2}{2\epsilon_0\hbar\Delta} \hspace{0.5cm}
\rm \xi_g=\frac{k_gN\mu_{12}^2}{2\epsilon_0\hbar\Delta}.
\end{equation}

\begin{figure}
\includegraphics[width=0.5\columnwidth]{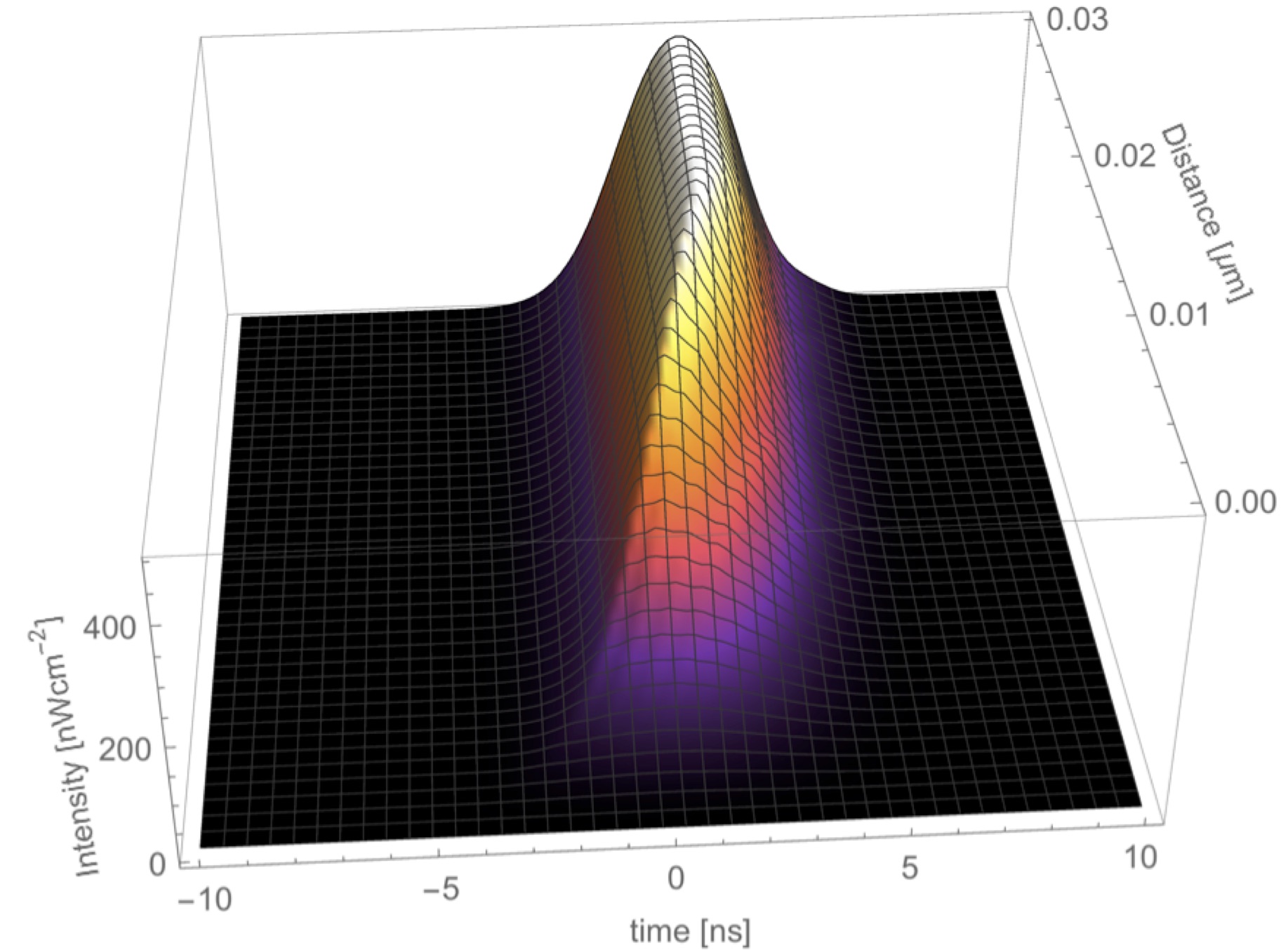}
\caption{\label{fig7} (Color online) Generated laser beam as a function of time and propagation distance. The data for Pump and Stokes laser were those of Fig.\,\ref{fig4}. The trigger laser was defined by $\rm \Omega_t=\Omega_{0t}Exp[-4\ln 2 \frac{t^2}{\tau_t^2}]$ with $\rm\Omega_{0t}=2\ns$ and $\rm \tau_t=7\,ns$. The photon energy of the trigger laser was set equal to $\rm E_{31}/2$.}
\end{figure}

These equations are valid as long as the trigger and generated beams are far detuned from any single-photon resonance of the system, and they do have a negligible effect on the medium so the population dynamics, and hence the coherence $\rm \rho_{13}(t)$, are not affected. Figure\,\ref{fig7} shows the generated laser beam with energy $\rm E_{31}/2$ as a function of time and propagation distance. The laser pulse grows rapidly, peaking at a distance for which the coherence is lost (see Fig.\,\ref{fig5}).

\begin{figure}
\includegraphics[width=0.5\columnwidth]{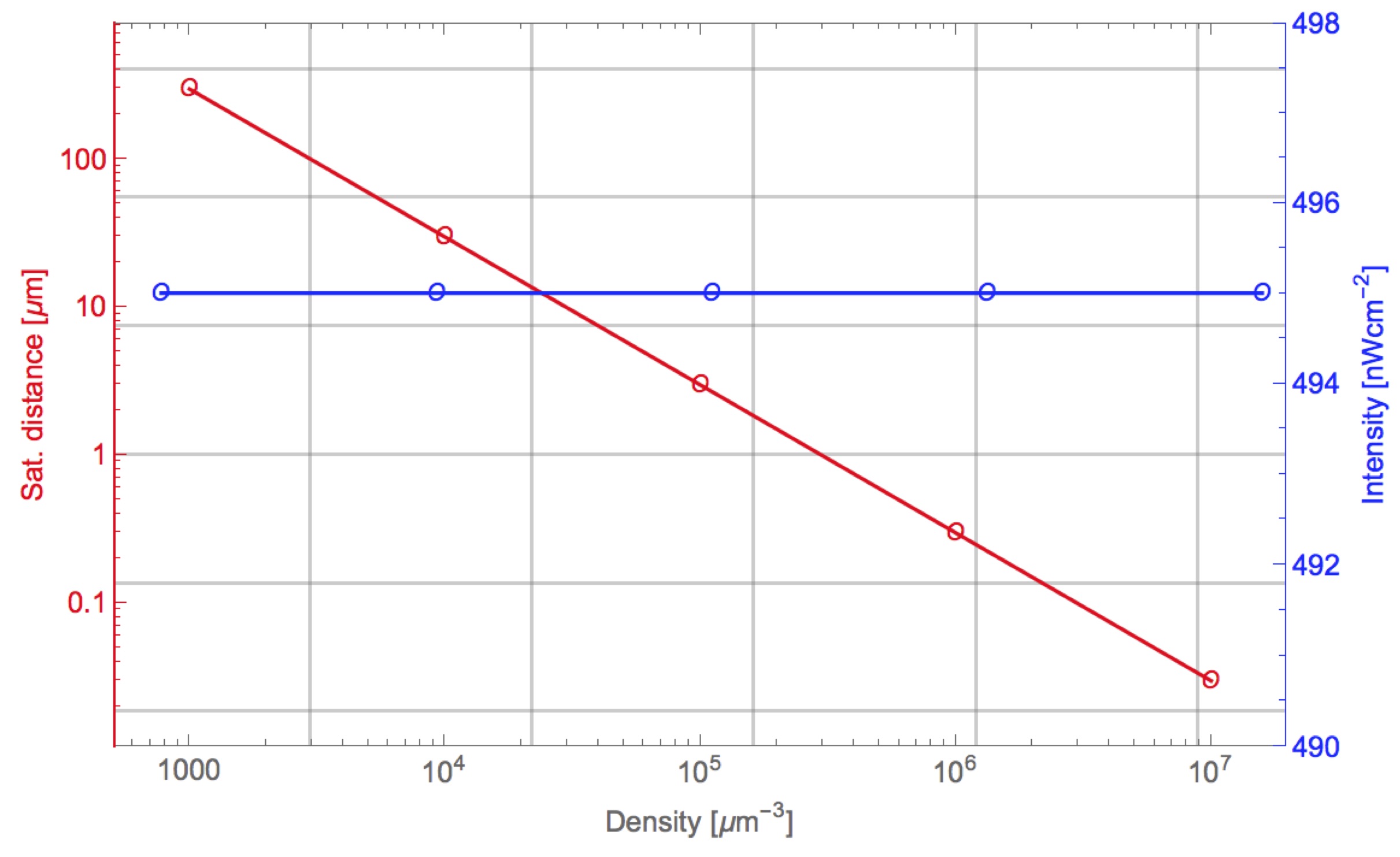}
\caption{\label{fig8} (Color online) Saturation distance and maximum intensity versus the density of the medium.}
\end{figure}

For a potential experimental implementation, it is useful to know the dependence of the saturation distance ---defined as the maximum propagation distance where coherence is induced in the system--- with respect to the density of the medium. Figure\,\ref{fig8} shows the outcome for different densities and a fit of them by
\begin{equation}
\label{satur}
\rm Sat.\,distance=\frac{296\cdot10^3}{N}[\mu m]
\end{equation}
being N the particle density in $\rm \mu m^{-3}$, together with the maximum intensity of the generated beam.

According to these results, we can conclude that the critical factor are the dimensions of the target where it is possible to induce a robust coherence, being this parameter related to the target density via a proportionality factor. Hence, the limiting factor for inducing coherence in a macroscopic target is the degradation of the laser pulses as they propagate. In this context, and bearing in mind the low probability of RENP, the manipulation of the laser-matter interaction (using for example shortcuts to adiabaticity \cite{Muga10}) may provide an alternative to overcome the limit imposed by Eq.\,\ref{satur}. Additionally, shortcuts to adiabaticity can be useful if decoherence terms play a relevant role in the dynamics of the system. Although two-photon CPR is robust against decoherence, in some situations it could be interesting to speed up the process. This is especially important for the nowadays situation because there is neither a definitive candidate for RENP nor a level scheme. So far there are different candidates but all of them present pros and cons \cite{Fukumi12}.

\subsection{Two dimensional numerical results}

Usually adiabatic techniques -as for example two-photon CPR- are pretty robust against variations in the experimental parameters, e.g., Rabi frequencies or laser detunings, being therefore ideal tools for the manipulation of atomic and molecular systems. Taking into account the inherent laser energy fluctuations and the spatial intensity distribution across the laser profile, it is clear that this robustness is mandatory for a successful experimental implementation. Since normally a realistic spatial laser profile is well described by a two dimensional Gaussian function, the overall effect is a proportionality factor ($\rm <1$) with respect to the one-dimensional simulations where a top-hat laser profile is implicitly assumed.

\begin{figure*}
\includegraphics[width=1\columnwidth]{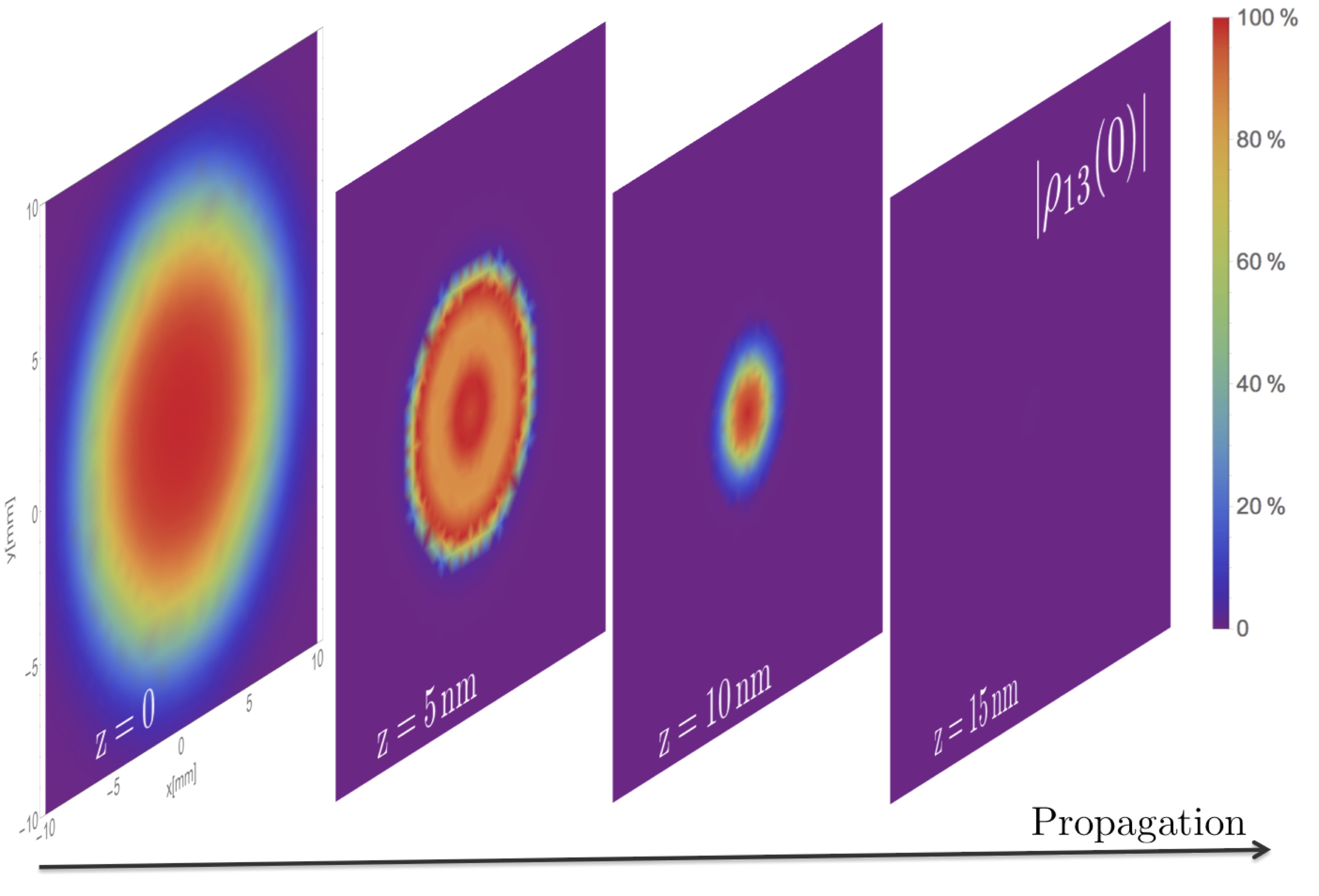}
\caption{\label{fig9} (Color online) Coherence $\rm |\rho_{13}(t)|$ for $\rm t=0$ across the spatial laser profile as a function of the propagation distance. The simulation parameters were those described in Fig.\,\ref{fig4}. We refer $\rm |\rho_{13}|=1/2$ to 100\%.}
\end{figure*}

Figure\,\ref{fig9} shows the variation of the maximum of the coherence $\rm |\rho_{13}(0)|$ across the spatial laser profile for the same set of parameters of Fig.\,\ref{fig4}. The spatial profile was described by a two dimensional Gaussian function with FWHM=10\,mm. According to the numerical results, for two-photon CPR there is not a heavy dependence on the laser intensity, making it possible to induce almost maximum coherence across the whole spatial laser profile.

\begin{figure*}
\includegraphics[width=1\columnwidth]{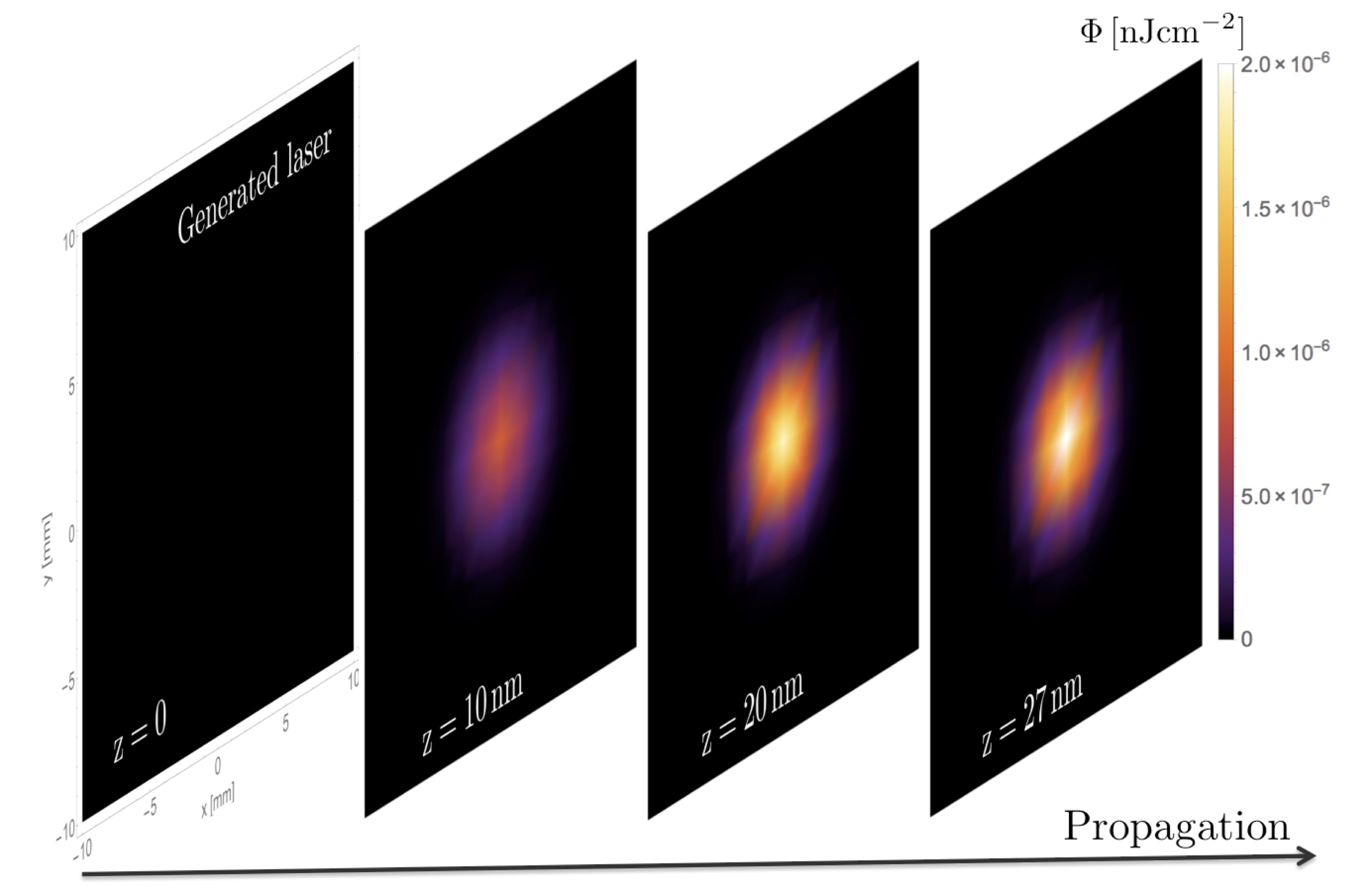}
\caption{\label{fig10} (Color online) Fluence of the externally triggered two-photon coherent emission.}
\end{figure*}

According to the level scheme shown in Fig.\,\ref{fig6}, Fig.\,\ref{fig10} shows spatial dependence of the coherent two-photon emission when the system is triggered by a laser with energy $\rm E_{31}/2$. The data used for the trigger laser were those of Fig.\,\ref{fig7} with a spatial profile defined by a two dimensional Gaussian function with FWHM=10\,mm. As we can clearly see, the coherent emission rapidly builds up in the system, saturating at the moment when coherence vanishes in the medium.

From these results, it is clear that the macrocoherence required for RENP becomes a source of background via two-photon coherent emission. The number of photons emitted per laser shot is $\rm \sim10^5$ a value which is several orders of magnitude larger than the number of photons expected in the RENP process \cite{Song16, Dinh13}. This represents a major experimental difficulty requiring high precision photon spectrometers capable not only of discriminating a small number of photons in a large background, but also possessing an excellent energy resolution.

\section{Conclusions}

In this manuscript we have described the propagation features, in one and two-dimensions, of two-photon CPR for the preparation of a macrocoherent state in a dense atomic or molecular medium. The numerical results show that although it is possible to induce maximum coherence in a relatively large volume, the interaction of the laser pulses with the medium degrades steadily the temporal profiles limiting the effective size of the medium. We have shown that it is possible to relate the maximum propagation distance where a coherence is established and the medium density.

Triggered by the proposal of RENP, we have also studied the coherent two-photon emission as a background source for the determination of the neutrino masses. Our results are sober. Although the construction of macrocoherence is an unavoidable step for the amplification of the extremely weak RENP signal, this macrocoherence amplifies the coherent two-photon emission in the system becoming, therefore, the main source of background and a very serious experimental drawback. In our opinion the detection of RENP is far from our actual technical capabilities. In this scenario we foresee two complementary lines of research for a future successful implementation of RENP. On the one hand, it seems mandatory to explore the implementation of RENP in systems where the RENP cross section is larger. In this respect, Yoshimura and Sasao have explored in a recent publication \cite{Yoshimura14}  a new type of RENP from a nucleus (or from inner core electrons) that could give rise to larger larger rates.

On the other hand, although adiabatic techniques (because of their stability against variations in the experimental parameters) offer the most promising tool for achieving large coherence in dense media, the degradation of the pulses as they propagate sets a limitation to the size of the medium and hence for RENP.  We must push this limit forward, maybe modifying the pulses characteristics in order that they become less sensitive to propagation, if we want to fulfill in a future the very demanding conditions ($\sim10^{21}$ atoms in a coherent state) for a possible experimental detection of RENP.

\section{Acknowledgments}

This work was supported by Ministerio de Econom\'ia y Competitividad (Spain) projects FIS2014-53371-C4-3-R and FIS2014-53371-C4-1-R.


\begin{thebibliography}{7}
\bibitem{Nota}
The three known neutrinos ($\nu_e$, $\nu_\mu$, $\nu_\tau$) are linear quantum superpositions of three massive states $\nu_i$ ($i=1,2,3$) with masses $m_i$, connected by a leptonic mixing matrix which can be parametrized as
\begin{equation}
\nonumber
  \label{eq:matrix}
  \rm U =
  \begin{pmatrix}
    \rm c_{12} c_{13}
    & \rm s_{12} c_{13}
    & 	\rm s_{13} e^{-i\delta_\text{CP}}
    \\
    \rm - s_{12} c_{23} - c_{12} s_{13} s_{23} e^{i\delta_\text{CP}}
    & \rm \hphantom{+} c_{12} c_{23} - s_{12} s_{13} s_{23}
    e^{i\delta_\text{CP}}
    &\rm  c_{13} s_{23}
    \\
    \rm \hphantom{+} s_{12} s_{23} - c_{12} s_{13} c_{23} e^{i\delta_\text{CP}}
    &\rm  - c_{12} s_{23} - s_{12} s_{13} c_{23} e^{i\delta_\text{CP}}
    & \rm c_{13} c_{23}
  \end{pmatrix} \begin{pmatrix}1 & 0 & 0 \\
0  & \rm e^{i\eta_1} & 0 \\
0  & 0 & \rm e^{i\eta_2} \end{pmatrix}
\end{equation}
where $\rm c_{ij} \equiv \cos\theta_{ij}$ and $\rm s_{ij} \equiv \sin\theta_{ij}$.  The phases $\eta_{i}$ are non-zero only if neutrinos are Majorana particles. If one chooses the convention where the angles $\theta_{ij}$ are taken to lie within the first quadrant, $\theta_{ij} \in [0, \pi/2]$,  and the CP phases $\delta_\text{CP},\eta_1,\eta_2 \in [0, 2\pi]$, then $\rm \Delta m^2_{21}=m_2^2-m_1^2>0$ by convention, and  $\rm \Delta m^2_{31}$ can be positive or negative. It is customary to refer to the first option as Normal Ordering (NO), and to the second one as Inverted Ordering (IO).
\end{thebibliography}
\end{document}